\documentclass[aps, prd, preprint, showpacs,superscriptaddress,groupedaddress]{revtex4-1}
\usepackage{amsmath,amsfonts,amsthm,amssymb}
\usepackage{bm}
\usepackage{graphicx}
\usepackage{enumerate,enumitem}
\usepackage{tikz,pgfplots}
\usepackage{placeins}

\pgfplotsset{compat=1.9}

\begin{document}

\title{Cosmological Models in Energy-Momentum-Squared Gravity}
\author{Charles V. R. Board}
\email{cvrb2@damtp.cam.ac.uk}
\author{John D. Barrow}
\email{J.D.Barrow@damtp.cam.ac.uk}

\begin{abstract}
We study the cosmological effects of adding terms of higher-order in the
usual energy-momentum tensor to the matter Lagrangian of general relativity.
This is in contrast to most studies of higher-order gravity which focus on
generalising the Einstein-Hilbert curvature contribution to the Lagrangian.
The resulting cosmological theories give rise to field equations of similar
form to several particular theories with different fundamental bases,
including bulk viscous cosmology, loop quantum gravity, $k$-essence, and
brane-world cosmologies. We find a range of exact solutions for isotropic
universes, discuss their behaviours with reference to the early- and
late-time evolution, accelerated expansion, and the occurrence or avoidance
of singularities. We briefly discuss extensions to anisotropic cosmologies
and delineate the situations where the higher-order matter terms will
dominate over anisotropies on approach to cosmological singularities.
\end{abstract}

\affiliation{DAMTP, Centre for Mathematical Sciences, University of Cambridge,
Wilberforce Rd., Cambridge CB3 0WA, U.K.}
\maketitle


\section{Introduction}

The twin challenges of naturally explaining two periods of accelerated
expansion during the history of the universe engage the attentions of many
contemporary cosmologists. The first period may have had a beginning and
necessarily came to an end when the universe was young and hot: it is called
a period of `inflation' and it leaves observable traces in the cosmic
microwave background radiation that are believed to have been detected. The
second period of acceleration began only a few billion years ago and is
observed in the Hubble flow traced by type IA supernovae; it is not known if
it will ever come to an end or is changing in any way. There are separate
non-unique mathematical descriptions of each of these periods of
acceleration but there is no single explanation of both of them, nor
any insight into whether or not they are related, or even random,
occurrences. For these reasons, there is continuing interest in all the
different ways in which expanding universes can undergo periods of
accelerated expansion. In the case of late-time acceleration the simplest
description of an effectively anti-gravitating stress, known as `dark
energy', is provided by introducing a cosmological constant ($\Lambda $)
into general relativity with a value arbitrarily chosen to match
observations.

The best-fit theory of this sort is dubbed $\Lambda$CDM and in its simplest
form is defined by six constants (which determine $\Lambda $) that can be
fixed by observation. One of those parameters is $\Lambda $ and its required
value is difficult to explain: it requires a theory that contributes an
effective vacuum stress of magnitude $\Lambda \sim (t_{pl}/t_{0})^{2}$ $\sim
10^{-120}$ at a time of observation $t_{0}\sim 10^{17}$s, where $t_{pl}\sim
10^{-43}$s is the Planck time \cite{bshaw}. Other descriptions that lead to
slowly evolving scalar fields in place of a constant $\Lambda $ have also
been explored, together with a range of modified gravity theories that
contribute anti-gravitating stresses. There are many such modifications and
extensions of Einstein's general relativity and they can be tuned to provide
acceleration at early or late times. So far, almost all of these
modifications to general relativity have focussed on generalising the
gravitational Lagrangian away from the linear function of the spacetime
curvature, $R$, responsible for the Einstein tensor in Einstein's equations.
A much-studied family of theories of this sort are those deriving from a
Lagrangian of the form $F(R)$, where $F$ is some analytic function. By
contrast, in this paper we will explore some of the consequences of
generalising the form of the matter Lagrangian in a nonlinear way, to some
analytic function of $T_{\mu\nu }T^{\mu\nu}$, where $T_{\mu\nu}$ is the
energy-momentum tensor of the matter stresses. This is more radical than
simply introducing new forms of fluid stress, like bulk viscosity or scalar
fields, into the Einstein equations in order to drive acceleration in Friedmann-Lema\^itre-Robertson-Walker (FLRW)
universes.

In \ref{sec:background} we discuss and motivate higher order contributions
to gravity from matter terms. In \ref{sec:frrtsection} we derive the
equations of motion for a generic $F(R,T_{\mu \nu }T^{\mu \nu })$
modification of the action with bare cosmological constant, before
specialising to the case $F(R,T_{\mu \nu }T^{\mu \nu })=R+\eta (T_{\mu \nu
}T^{\mu \nu })^{n}$. We then investigate several features of the isotropic
cosmology in this theory in \ref{sec:isocossec} and, finally, move to the
anisotropic Bianchi type I setting in \ref{sec:anisocossec}.

\section{Background}

\label{sec:background}

\subsection{Field equations}

Einstein's theory of general relativity (GR) with cosmological constant $%
\Lambda $ can be derived from the variation of the action, 
\begin{equation}
S=\frac{1}{2\kappa }\int \sqrt{-g}(R-2\Lambda )d^{4}x+\int \sqrt{-g}%
L_{m}d^{4}x,  \label{EHac}
\end{equation}
where $\kappa =8\pi G$ and $L_{m}$ is the matter Lagrangian, which we will
take to describe a perfect fluid; $R\equiv R_{a}^{a}$, where $R_{b}^{a}$ is
the Ricci tensor, and $g$ is the determinant of the metric itself. Here, and
in all that follows, we use units in which $c=1$.

An isotropic and homogeneous universe may be described by the FLRW metric: 
\begin{equation}
ds^{2}=-dt^{2}+a^{2}(t)\left( \frac{dr^{2}}{1-kr^{2}}+r^{2}\left( d\theta
^{2}+\sin ^{2}\theta d\phi ^{2}\right) \right) ,  \label{metric}
\end{equation}
where $k$, the curvature parameter, takes the values \{$-1$, $0$, $+1$\}
corresponding to open, flat and closed 3-spaces, respectively; $t$ is the
comoving proper time and $a(t)$ is the expansion scale factor.

There are many proposals to modify or extend the $\Lambda $CDM cosmological
picture. These fall broadly into two categories, depending on which side of
the Einstein field equations is modified. We can modify the right-hand side
of the Einstein equations by adding new forms of matter that will drive
expansion either at early times, as in the theory of inflation, or at late
times, such as in quintessence or $k$-essence scenarios \cite{dyndarkeng}.
Alternatively, we can modify the left-hand side of the Einstein equations in
order to modify the effect of gravity itself. There are several ways to do
this, including $F(R)$ theories \cite{frtheories} in which the Ricci scalar
in \eqref{EHac} is replaced by some function $f(R)$, so-called $F(T)$
theories in which we modify the teleparallel equivalent of general
relativity \cite{ftbarrow}, or scalar-tensor theories in which a scalar
field is coupled to the Ricci scalar.

\subsection{Higher-order matter contributions}

The type of generalisation of general relativity we will explore in this
paper looks to include higher-order contributions to the right-hand side of
the Einstein equations, where the material stresses appear. This results in
field equations that include new terms that enter at high densities and
pressures, which may be anti-gravitational in their effects. Typically, they
affect the cosmological model at high densities and may alter the
conclusions regarding the appearance of spacetime singularities in the
finite cosmological past. Conversely, we might expect their effects at late
times and low cosmological densities to be very small. Even within general
relativity, there is scope to include high-order matter contributions, as
the Einstein equations have almost no content unless some prescription or
constraint is given on the forms of matter stress. Thus, in the
general-relativistic Friedmann models, we can introduce non-linear stresses
defined by relations between pressure, $p$, and density, $\rho $, of the
form $\rho +p=\gamma \rho ^{n},$ \cite{jbgen}, or $f(\rho )$, \cite{odint},
where $\gamma \geq 0$ and $n$ are constants, or include a bulk viscous stress
into the equation of state of the standard form $p=(\gamma -1)\rho
-3H\varsigma (\rho ),$where $H$ is the Hubble expansion rate and $\varsigma
\geq 0$ is the bulk viscosity coefficient \cite{bulk}. The so-called Chaplygin and generalised Chaplygin gases are just special cases of these
bulk viscous models, and choices of $n$ or $\varsigma \propto \rho ^{m}$
introduce higher-order matter corrections. Similarly, the choice of
self-interaction potential $V(\phi )$ for a scalar field can also introduce
higher-order matter effects into cosmology. Analogously, in scalar-tensor
theories like Brans-Dicke (BD) which are defined by a constant BD coupling
constant, $\omega $, generalisations are possible to the cases where $\omega 
$ becomes a function of the BD scalar field. In all these extensions of the
standard relativistic perfect fluid cosmology there will be several critical
observational tests which will constrain them. In particular, in
higher-order matter theories the inevitable deviations that can occur from
the standard thermal history in the early radiation era will change the
predicted abundances of helium-4 and deuterium and alter the detailed
structure of the microwave background power spectrum. Also, as we studied
for Brans-Dicke theory \cite{lmb, chen}, changes in the cold dark matter
dominated era evolution can shift the time when matter and radiation
densities are equal. This is the epoch when matter perturbations begin to
grow and sensitively determines the peak of the matter power spectrum. At a
later nonlinear stage of the evolution, higher-order gravity theories will
effect the formation of galactic halos. This has been investigated for bulk
viscous cosmologies by Li and Barrow \cite{Li}. These observational
constraints will form the subject of a further paper and will not be
discussed here.

If we depart from general relativity, then various simple quantum
gravitational corrections are possible, and have been explored. The most
well known are the loop quantum gravity (LQG) \cite{LQG} and brane-world 
\cite{bw} scenarios that contribute new quadratic terms to the Friedmann
equation for isotropic cosmologies by replacing $\rho $ by $\rho (1\pm
O(\rho ^{2}))$ in the Friedmann equation, where the $-$ contribution is from
LQG and the $+$ is from brane-world scenarios. The impact on anisotropic cosmological
models is more complicated and not straightforward to calculate \cite{singh,
gupt}. In particular, we find that simple forms of anisotropic stress are no
longer equivalent to a $p=\rho $ fluid as we are used to finding in general
relativity. Our study will be of a type of higher-order matter corrections
which modify the Friedmann equations in ways that include both of the
aforementioned types of phenomenological modification to the form of the
Friedmann equations, although the underlying physical theory does not
incorporate the LQC or brane-world models or reduce to them in a limiting
case.

Standard $F(R)$ theories of gravity \cite{frtheories} can be generalised to
include a dependence of the form

\begin{equation}
S=\frac{1}{2\kappa }\int \sqrt{-g}F(R,L_{m})d^{4}x.
\end{equation}

This is in some sense an extremal extension of the Einstein-Hilbert action,
as discussed in \cite{frlm}. If the coupling between matter and gravity is
non-minimal, then there will be an extra force exerted on matter, resulting
in non-geodesic motion and a violation of the equivalence principle. This
type of modification has been investigated in several contexts, particularly
when the additional dependence on the matter Lagrangian arises from $F$
taking the form $F(R,\mathcal{T})$ where $\mathcal{T}$ is the trace of the
energy-momentum tensor \cite{frtod}.

A theory, closely related to $F(R,\mathcal{T})$ gravity, that allows the
gravitational Lagrangian to depend on a more complicated scalar formed from
the energy-momentum tensor is provided by $F(R,\mathbf{T}^{2})$, where $%
\mathbf{T^{2}}\equiv T_{\mu \nu }T^{\mu \nu }$ is the scalar formed from the
square of the energy-momentum tensor. This was first discussed in \cite{frtt}%
, and the special case with 
\begin{equation}
F(R,\mathbf{T^{2}})=R+\eta \mathbf{T^{2},}  \label{emsgF}
\end{equation}%
where $\eta $ is a constant, was also discussed in \cite{emsg}, where the
authors investigated the possibility of a bounce at early times when $\eta <0
$ (although in that paper they used the opposite sign convention to us for $%
\eta $), and also found an exact solution for charged black holes in the
extended theory. In \cite{Liu:2016qfx} a similar form, with additional cross
terms between the Ricci and Energy-momentum tensors, was discussed as
arising from quantum fluctuations of the metric tensor. Recently the authors of \cite%
{Akarsu:2017ohj} investigated the late time acceleration of universes
described by this model in the dust-only case, and used observations
of the Hubble parameter to constrain the parameters of the theory.

We would expect the theory derived from \eqref{emsgF} to provide different
physics to the $F(R,\mathcal{T})$ case. Indeed, one example of this is the
case of a perfect fluid with equation of state $p=-\frac{1}{3}\rho $. The
additional terms in $F(R,\mathcal{T})$ will vanish as $\mathcal{T}=0$, but
in the $F(R,\mathbf{T^{2}})$ theory the extra terms in $\mathbf{T^{2}}$ will
not vanish and we will find new cosmological behaviour. In \ref%
{sec:frrtsection}, we will investigate the cosmological solutions in a more
general setting, where the $\mathbf{T^{2}}$ term may be raised to an
arbitrary power.

\section{Field Equations for $F(R,T_{\protect\mu \protect\nu }T^{\protect\mu 
\protect\nu })$ Gravity with Cosmological Constant}

\label{sec:frrtsection}

In \cite{emsg} the Friedmann equations were derived in the case where $F$ is
given by \eqref{emsgF}, for a flat FLRW cosmology. A `bare' cosmological
constant was also included on the left-hand side of the field equations
(rather than as an effective energy-momentum tensor for the vacuum). In \cite%
{frtt}, the field equations were derived without a cosmological constant and
specialised to two particular models. We first derive the equations of
motion with a cosmological constant for general $F$, before specialising to
theories where the additional term takes the form $(\mathbf{T^{2})}^{n}$,
and determining the FLRW equations with general curvature. In GR, the
cosmological constant can be considered to be, equivalently, either a `bare'
constant on the left-hand side of the Einstein equations, or part of the
matter Lagrangian. As discussed in \cite{emsg}, the two are no longer
equivalent in this theory, due to the non-minimal nature of the
curvature-matter couplings. A similar inequivalence also occurs in other
models that introduce non-linear matter terms, including loop quantum
cosmology. We will assume that the cosmological constant arises in its bare
form as part of the gravitational action. This gives the modified action

\begin{equation}
S=\frac{1}{2\kappa }\int \sqrt{-g}(F(R,T^{\mu \nu }T_{\mu \nu })-2\Lambda )\
d^{4}x+\int L_{m}\sqrt{-g}\ d^{4}x,  \label{action}
\end{equation}

where $\ L_{m}$ is taken to be the same as the matter component contributed
by $T_{\mu \nu }$. Since the gravitational Lagrangian now depends on $%
\mathbf{T^{2}}$, we note that the new terms in the variation of the action
will arise from the variation of this square, via $\delta (T_{\mu \nu
}T^{\mu \nu })$. To calculate this, we define $T_{\mu \nu }$ by

\begin{equation}
T_{\mu\nu}=-\frac{2}{\sqrt{-g}}\frac{\delta(\sqrt{-g}L_m)}{\delta g^{\mu\nu}}%
.
\end{equation}

We enforce the condition that $L_{m}$ depends only on the metric components,
and not on their derivatives, to find 
\begin{equation}
T_{\mu \nu }=g_{\mu \nu }L_{m}-2\frac{\partial L_{m}}{\partial g^{\mu \nu }}.
\end{equation}

Varying with respect to the inverse metric, we define 
\begin{equation}
\theta _{\mu \nu }\equiv \frac{\delta (T_{\alpha \beta }T^{\alpha \beta })}{
\delta g^{\mu \nu }}=-2L_{m}(T_{\mu \nu }-\frac{1}{2}g_{\mu \nu }T)-TT_{\mu
\nu }+2T_{\mu }^{\alpha }T_{\nu \alpha }-4T^{\alpha \beta }\frac{\partial
^{2}L_{m}}{\partial g^{\mu \nu }\partial g^{\alpha \beta }},  \label{thetaT}
\end{equation}
where $T$ is the trace of the energy-momentum tensor. Varying the action in
this way, we find 
\begin{equation}
\delta S=\frac{1}{2\kappa }\int \{F_{R}\delta R+F_{T^{2}}\delta (T_{\mu \nu
}T^{\mu \nu })-\frac{1}{2}g_{\mu \nu }F\delta g^{\mu \nu }+\Lambda +\frac{1}{
\sqrt{-g}}\delta (\sqrt{-g}L_{m})\}d^{4}x,
\end{equation}
where subscripts denote differentiation with respect to $R$ and $\mathbf{\
T^{2}}$, respectively.

From this variation we obtain the field equations:

\begin{equation}  \label{frttfield}
F_RR_{\mu\nu}-\frac{1}{2}Fg_{\mu\nu}+\Lambda
g_{\mu\nu}+(g_{\mu\nu}\nabla_\alpha\nabla^\alpha-\nabla_\mu\nabla_\nu)F_R=
\kappa(T_{\mu\nu}-\frac{1}{\kappa}F_{\mathbf{T^2}}\theta_{\mu\nu}).
\end{equation}

These reduce, as expected, to the field equations for $F(R)$ gravity in the
special case where $F(R,\mathbf{T^{2}})=F(R)$ \cite{frtheories} and to the
Einstein equations with a cosmological constant when $F(R,\mathbf{T^{2}})=R$.

We will assume that the matter component can be described by a perfect
fluid, 
\begin{equation}
T_{\mu \nu }=(\rho +p)u_{\mu }u_{\nu }+pg_{\mu \nu },  \label{perfectfluid}
\end{equation}
where $\rho $ is the energy density and $p$ the pressure; hence

\begin{equation}
T_{\mu\nu}T^{\mu\nu}=\rho^2+3p^2.
\end{equation}

Furthermore, we take the Lagrangian $L_{m}=p$. This means that the final
term in the definition of $\theta _{\mu \nu }$ vanishes and allows us to
calculate the form of $\theta _{\mu \nu }$ independently of the function $F$%
. Substituting \eqref{perfectfluid} into \eqref{thetaT}, we find

\begin{equation}
\theta_{\mu\nu}=-(\rho^2+4p\rho+3p^2)u_\mu u_\nu.
\end{equation}

We now proceed to specify a particular form for $F(R,\mathbf{T^{2}})$ which
includes and generalises the models used in \cite{frtt} and for
energy-momentum-squared gravity in \cite{emsg} (EMSG). This form is 
\begin{equation}
F(R,T_{\mu \nu }T^{\mu \nu })=R+\eta (T_{\mu \nu }T^{\mu \nu })^{n},
\end{equation}

where $n$ need not be an integer. This corresponds to EMSG in the case $n=1$%
, and to Models A and B of \cite{frtt} when $n=1/2$ and $n=1/4,$
respectively; it reduces the field equations to

\begin{equation}
R_{\mu\nu}-\frac{1}{2}Rg_{\mu\nu}+\Lambda g_{\mu\nu}=\kappa(T_{\mu\nu}+\frac{
\eta}{\kappa}(T_{\alpha\beta}T^{\alpha\beta})^{n-1}\left[\frac{1}{2}
(T_{\alpha\beta}T^{\alpha\beta})g_{\mu\nu}-n\theta_{\mu\nu}\right]),
\end{equation}

which we rewrite as

\begin{equation}
G_{\mu\nu}+\Lambda g_{\mu\nu}=\kappa T^{\text{eff}}_{\mu\nu},
\end{equation}

where $G_{\mu \nu }$ is the Einstein tensor, to show the relationship to
general relativity. Continuing with the perfect fluid form of the
energy-momentum tensor, this expands to give: 
\begin{equation}
\begin{split}
G_{\mu \nu }+\Lambda g_{\mu \nu }=\kappa ((\rho & +p)u_{\mu }u_{\nu
}+pg_{\mu \nu }) \\
& +\eta (\rho ^{2}+3p^{2})^{n-1}\left[\frac{1}{2}(\rho ^{2}+3p^{2})g_{\mu
\nu }+n(\rho +p)(\rho +3p)u_{\mu }u_{\nu }\right].
\end{split}
\label{fieldeq}
\end{equation}

\section{Isotropic Cosmology}

\label{sec:isocossec} If we assume a FLRW universe with curvature parameter $k$, we find the generalised Friedmann equation, 
\begin{equation}
\left( \frac{\dot{a}}{a}\right) ^{2}+\frac{k}{a^{2}}=\frac{\Lambda }{3}
+\kappa \frac{\rho }{3}+\frac{\eta }{3}(\rho ^{2}+3p^{2})^{n-1}\left[(n- 
\frac{1}{2})(\rho ^{2}+3p^{2})+4n\rho p\right] ,  \label{ttfriedmann}
\end{equation}

and acceleration equation 
\begin{equation}
\frac{\ddot{a}}{a}=-\kappa \frac{\rho +3p}{6}+\frac{\Lambda }{3}-\frac{\eta 
}{3}(\rho ^{2}+3p^{2})^{n-1}\left[ \frac{n+1}{2}(\rho ^{2}+3p^{2})+2n\rho p%
\right] .  \label{frttaccel}
\end{equation}

If the matter field obeys a barotropic equation of state, $p=w\rho $ with $w$
constant, then the non-GR terms are all of the form $\rho ^{2n}$ multiplied
by a constant. Thus, the generalised Friedmann equation becomes

\begin{equation}
\left( \frac{\dot{a}}{a}\right) ^{2}+\frac{k}{a^{2}}=\frac{\Lambda }{3}%
+\kappa \frac{\rho }{3}+\frac{\eta \rho ^{2n}}{3}A(n,w),
\label{frttfriedmannb}
\end{equation}
where $A$ is a constant depending on the choice of $n$ and $w$, given by 
\begin{equation}
A(n,w)\equiv (1+3w^{2})^{n-1}\left[(n-\frac{1}{2})(1+3w^{2})+4nw\right],
\end{equation}
and the acceleration equation becomes

\begin{equation}
\frac{\ddot{a}}{a}=-\kappa \frac{1+3w}{6}\rho +\frac{\Lambda }{3}-\frac{\eta
\rho ^{2n}}{3}B(n,w),  \label{frttaccelb}
\end{equation}
where $B$ a constant given by 
\begin{equation}
B(n,w)\equiv (1+3w^{2})^{n-1}\left[\frac{n+1}{2}(1+3w^{2})+2nw\right].
\end{equation}

Finally, we determine the generalised continuity equation, by
differentiating the generalised Friedmann equation,

\begin{equation}
\dot{\rho}=-3\frac{\dot{a}}{a}\rho(1+w)\left[\frac{\kappa+\eta\rho^{2n-1}n(1+3w)(1+3w^2)^{n-1}}{\kappa+2\eta\rho^{2n-1}nA(n,w)}\right] ,  \label{frttcontb}
\end{equation}
where we have written it in a form that makes clear the generalisation of
the GR case.

We can see immediately that there is an interesting difference between the
FLRW equations in GR and in EMSG. When $\eta =0$ there are solutions with
finite $a,\dot{a},$and $\rho $ but infinite values of $p$ and $\ddot{a}$.
These are called sudden singularities \cite{sudd1, sudd2, sudd3} and can be
constructed explicitly. In EMSG, where $\eta \neq 0$, the appearance of the
pressure, $p$, explicitly in the Friedmann equation changes the structure of
the equations and the same type of sudden singularity is no longer possible
at this order in derivatives of $a$.

\subsection{Integrating the continuity equation}

We now attempt to determine the cosmological behaviour of some cases where
the modified continuity equation can be integrated exactly. We find four
simply integrable cases: two of these are for fixed $w$ independent of the
value of $n$, the other two occur for specific values of $w$ dependent on
the choice of $n$, although we note that some of these integrable cases may
coincide, depending on our choice of the exponent, $n$.

The first case that can be integrated is for the equation of state
corresponding to dark energy, $w=-1$, where the entire right-hand side of %
\eqref{frttcontb} vanishes, and so $\rho \equiv \rho _{0}$, a constant. In
this case we expect to find a solution to the modified Friedmann equation
that is the same as the solution in GR except with altered constants, which
results in a de Sitter solution where $H\equiv \frac{\dot{a}}{a}=$ constant,
and the universe expands exponentially.

Next, we consider the case $w=-\frac{1}{3}$, which corresponds to an
effective perfect fluid representing a negative curvature, so the numerator
in the modified continuity equation becomes simply $\kappa ,$ and we can
integrate \eqref{frttcontb} since 
\begin{align}
\dot{\rho}\left(\frac{1}{\rho}+\frac{2\eta n A(n,\frac{-1}{3})}{\kappa}\rho^{2n-2}\right)=&-2\frac{\dot{a}}{a}, \\
\frac{d}{dt}\left(\ln(\rho)-\frac{\eta n (\frac{4}{3})^n}{(2n-1)\kappa}\rho^{2n-1}\right)=&\frac{d}{dt}(\ln(a^{-2})), \\
\rho\exp\left(-\frac{\eta n (\frac{4}{3})^n}{(2n-1)\kappa}\rho^{2n-1}\right)=&Ca^{-2},
\end{align}
with $C>0$ a constant of integration.

We can also integrate the continuity equation when the correction factor in %
\eqref{frttcontb} is equal to $1$, which occurs when 
\begin{equation}
(1+3w)(1+3w^2)^{n-1}=2A(n,w).  \label{c3}
\end{equation}
The continuity equation then reduces to the standard GR form for these
special values, $w=w_{\ast }$, and so we have 
\begin{equation}
\rho =Ca^{-3(1+w_{\ast })}.
\end{equation}

The final possibility that we consider is when 
\begin{equation}
n(1+3w)(1+3w^2)^{n-1}=A(n,w),  \label{c4}
\end{equation}
in which case we can write \eqref{frttcontb} as 
\begin{equation}
\frac{d}{dt}\left(\ln(\kappa\rho+n\eta\rho^{2n}(1+3w_{*})(1+3w_{*}^2)^{n-1})\right)=\frac{d}{dt}(\ln(a^{-3(1+w_{*})})),
\end{equation}
which integrates to

\begin{equation}
\kappa\rho+n\eta\rho^{2n}(1+3w_{*})(1+3w_{*}^2)^{n-1}=Ca^{-3(1+w_{*})}.
\end{equation}

We note that, depending on the choice of exponent $n$, some of the second
pair of solutions may exist for multiple choices of $w$, or may coincide
with each other, or with the $w=-1$, $w=-\frac{1}{3}$ cases. Also, for some
choices of $n,$ there may be no solutions at all.

Finally, note that only one of these solutions allows easy integration of
the modified Friedmann equation \eqref{frttfriedmannb}. This is the case
when $w=-1$ and so $\rho =\rho _{0}$. In this case the Friedmann-like
equation becomes 
\begin{equation}
\left( \frac{\dot{a}}{a}\right) ^{2}+\frac{k}{a^{2}}=\alpha (\Lambda ,n),
\end{equation}
where $\alpha $ is a constant given by 
\begin{equation}
\alpha (\Lambda ,n)\equiv \frac{\Lambda }{3}+\kappa \frac{\rho _{0}}{3}- 
\frac{\eta \rho _{0}^{2n}}{6}4^{n}.
\end{equation}

The solution to the modified Friedmann equation is then given by 
\begin{equation}
a(t)=\frac{1}{2\sqrt{\alpha }}(C\sqrt{\alpha }+\frac{k}{C\sqrt{\alpha }}
)\cosh (\sqrt{\alpha }t)\pm (C\sqrt{\alpha }-\frac{k}{C\sqrt{\alpha }})\sinh
(\sqrt{\alpha }t)
\end{equation}
where $C$ is a new constant of integration. Equivalently, we can write this
solution in terms of exponentials as 
\begin{equation}
a(t)=\frac{1}{2\sqrt{\alpha }}\left( C\sqrt{\alpha }e^{\sqrt{\alpha }t}+%
\frac{k}{C\sqrt{\alpha }}e^{-\sqrt{\alpha }t}\right)
\end{equation}
as well as its time reversal, $t\rightarrow -t$. Assuming $\alpha >0$, we
can see that this reduces to the expected de Sitter solution from general
relativity in the case $k=0$, as we would expect. If $\alpha <0$ then,
writing instead $\alpha \rightarrow -\alpha $, there is a real solution only
for negative curvature, where we must choose $k=-C^{2}\alpha $, giving the
anti-de Sitter solution 
\begin{equation}
a(t)=C\cos (\sqrt{\alpha }t).
\end{equation}
It is important to note that because of the form of $\alpha $, unlike in the
unmodified case, we do not necessarily require a negative cosmological
constant to find this solution. We would expect this anti-de Sitter analogue
to appear whenever $\eta >0$, for suitable choices of $\rho _{0}$ and $n$.

This solution is very similar to the case of $w=-1$ in GR, where we can
rewrite the cosmological constant as a perfect fluid with this equation of
state. This is possible in GR because the continuity equations for
non-interacting multi-component fluids decouple, allowing us to treat them
independently. Unfortunately, because of the additional non-linear terms
arising in these $F(R,\mathbf{T^{2}})$ models (except in the special case $%
n=1/2$), we cannot decouple different fluids in this way and then
subsequently superpose them in our Friedmann-like equations. This means that
we cannot replace the curvature or cosmological constant terms with perfect
fluids with $w=-1/3$ and $-1$ as in classical GR. However, for some choices
of $n$ and $\eta ,$ the correction terms can themselves provide an
additional late-time or early inflationary repulsive force, removing the
need for an explicit cosmological constant.

\subsection{Energy-momentum-squared gravity: the case $n=1$}

If we fix our choice of $n$, then we can say more about the behaviour of the
specific solutions that arise. In what follows we consider primarily the
case $n=1$ which was originally discussed in \cite{emsg}, under the name
`energy-momentum squared gravity'. After specialising to $n=1$, we can say
more about the solutions to the continuity equation found in the previous
section, and investigate the modified Friedmann equations. The form of the
Friedmann equations, after setting $n=1$ in \eqref{frttfriedmannb}, %
\eqref{frttaccelb} and \eqref{frttcontb}, is: 
\begin{align}
\left( \frac{\dot{a}}{a}\right) ^{2}+\frac{k}{a^{2}}=& \frac{\Lambda }{3}
+\kappa \frac{\rho }{3}+\frac{\eta \rho ^{2}}{6}(3w^{2}+8w+1)
\label{EMSGFriedmann} \\
\frac{\ddot{a}}{a}=& \frac{\Lambda }{3}-\kappa \frac{1+3w}{6}\rho -\frac{
\eta \rho ^{2}}{3}(3w^{2}+2w+1) \\
\dot{\rho}=& -3\frac{\dot{a}}{a}\rho (1+w)\frac{\kappa +\eta \rho (1+3w)}{
\kappa +\eta \rho (3w^{2}+8w+1)}
\end{align}

The new terms in the Friedmann equations are quadratic in the energy
density, which we would expect to dominate in the very early universe as $%
\rho \rightarrow \infty $. Additionally, if we choose $\eta <0,$ then the
modified Friedmann equations in this model are similar to the effective
Friedmann equations arising in loop quantum cosmology, \cite{LQG}, where 
\begin{equation}
\left( \frac{\dot{a}}{a}\right) =\frac{\kappa }{3}\rho \left( 1-\frac{\rho }{
\rho _{\text{crit}}}\right) ,
\end{equation}
which may warrant further investigation. An analogous higher-order effect
occurs in brane world cosmologies, where there is an effective equation of
state with \cite{bher, bmar, coley, coley2}

\begin{equation}
p^{eff}=\frac{1}{2\lambda }(\rho ^{2}+2p\rho );\lambda >0\text{ constant.}
\end{equation}

We briefly summarise the values of $w$ for which the results of the previous
section allow us to integrate the Friedmann equation and find the values of $%
w$ that satisfy \eqref{c3} and \eqref{c4}. If we set $n=1$ then %
\eqref{c3} reduces to 
\begin{equation}
3w^{2}+5w=0,
\end{equation}%
which has solutions $w=-\frac{5}{3}$ and $w=0$. The $w=0$ solution describes
`dust' matter. The case $w=-5/3$ corresponds to some form of phantom energy,
which will result in a Big Rip singularity, \cite{cald}, at finite future
time.

Alternatively, solving \eqref{c4} for $n=1$ gives 
\begin{equation}
3w^{2}+2w-1=0,
\end{equation}%
which has solutions $w=-1$ and $w=\frac{1}{3}$. The first of these has
already been found for all $n$ as the first case above, whilst the second
gives a solution corresponding to blackbody radiation. Hence, we have exact
solutions to the continuity equation for the cases $w=\{-\frac{5}{3},-1,-%
\frac{1}{3},0,\frac{1}{3}\}$ which include the physically important cases of
dust and radiation.

The equation of state $p=0$ corresponds to pressureless dust or
non-relativistic cold dark matter, and as shown above, we recover the same
dependence of the energy density on the scale factor as in the GR case, 
\begin{equation}
\rho =Ca^{-3}.  \label{cons}
\end{equation}

If we combine this with the modified acceleration and Friedmann equations
for $w=0$ we find 
\begin{equation}
a\ddot{a}+2\dot{a}^{2}+k=\frac{\Lambda }{2}a^{2}+\frac{\kappa }{4C}a^{-1}.
\end{equation}

If we consider only flat space ($k=0$) then we find 
\begin{equation}
a(t)=(4\Lambda )^{-\frac{1}{3}}((C^{2}+D+1)\cosh \left( \sqrt{\frac{3\Lambda 
}{2}}t\right) +(C^{2}+D-1)\sinh \left( \sqrt{\frac{3\Lambda }{2}}t\right)
-2C)^{\frac{1}{3}},
\end{equation}

where $D$ is a constant of integration, and we have eliminated a further
constant by a covariant translation of the time coordinate. We can then find 
$\rho $ explicitly, using \eqref{cons}. We can see, however, that this form
of the solution does not capture the case $\Lambda =0$. In this case,
instead we find the solution 
\begin{equation}
a(t)=\left( \frac{3}{8C}\right) ^{\frac{1}{3}}(C^{2}t^{2}-16D)^{\frac{1}{3}},
\end{equation}

which  gives the GR dust behaviour of $a\sim t^{\frac{2}{3}}$ at large $t$.

In the case of $w=-\frac{1}{3}$, we can write

\begin{equation}
\rho \exp \left( -\frac{4\eta }{3\kappa }\rho \right) =Ca^{-2}.
\end{equation}%
After differentiation and multiplication by $a^{2}$, we can write 
\begin{equation}
\frac{\dot{a}}{a}=\frac{\dot{\rho}}{\rho }(1-\frac{4\eta }{3\kappa }\rho )
\end{equation}%
and so in the case $k=0$ we can write the Friedmann equation in terms of $%
\rho $ without any exponentials, as

\begin{equation}
\left( \frac{\dot{\rho}}{\rho }\right) ^{2}(1-\frac{4\eta }{3\kappa }\rho
)^{2}\frac{1}{4C^{2}}=\frac{\Lambda }{3}+\frac{\kappa }{3}\rho -\frac{2\eta 
}{9}\rho ^{2}.
\end{equation}%
Finally, in the case of $w=\frac{1}{3}$, which corresponds to radiation, 
\cite{emsg} gave a solution in the case of flat space, $a(t)\propto \sqrt{%
\cosh (\alpha t)}$ where $\alpha \equiv \sqrt{\frac{4\Lambda }{3}}$. We see
that in this case we can write the continuity equation as

\begin{equation}
\kappa \rho +2\eta \rho ^{2}=Ca^{-4},
\end{equation}%
and that in the Friedmann and acceleration equations, the density terms are
of equal magnitude but opposite sign. We can then sum the two to find our
equation for $a(t)$ 
\begin{equation}
\left( \frac{\dot{a}}{a}\right) ^{2}+\frac{\ddot{a}}{a}+\frac{k}{a^{2}}=%
\frac{2\Lambda }{3},
\end{equation}

which we solve by use of the substitution $y=a^{2}$ to find 
\begin{equation}
a^{2}(t)=\frac{1}{4\Lambda }((1+9k^{2}-12\Lambda D)\cosh (\alpha
t)+(1-9k^{2}+12\Lambda D)\sinh (\alpha t)+6k)
\end{equation}%
for all $\Lambda $, $k$ non-zero, with $\alpha \equiv \sqrt{\frac{4\Lambda }{%
3}},$ as above. In the $\Lambda =0$ subcase, we find the solutions

\begin{equation}
a^{2}(t)=\left\{ 
\begin{array}{cc}
Dt-kt^{2} & k\neq 0 \\ 
Dt & k=0 \\ 
& 
\end{array}%
\right. .
\end{equation}

\subsection{de Sitter-like solutions}

de Sitter solutions arise in EMSG theory. They have constant density and
Hubble parameter, which includes the case $w=-1$. In $\Lambda $CDM we expect
this to arise in two situations. The first is when we have $\rho \equiv 0$,
that is an empty universe whose expansion is controlled solely by $\Lambda ,$
and the second is the similar dark-energy equation of state $w=-1$ for which
the perfect fluid behaves as a cosmological constant. In EMSG we find that there
is an extra family of de Sitter solutions. We describe them first for
general $n$, then specialise to EMSG.

Since we are searching for solutions with $H\equiv H_{0}$ and $\rho \equiv
\rho _{0}$, from \eqref{frttfriedmannb} we must have $k=0$, and the
Friedmann equation then reduces to an algebraic one for $H^{2}$ in terms of $%
\rho _{0}$. Similarly, since $\dot{H}=0$, \eqref{frttaccelb} reduces to
another relation for $H^{2}$. Equating the two to remove $H^{2}$ and
simplifying, we find that $\rho _{0}$ must satisfy 
\begin{equation}
\rho _{0}(1+w)(\kappa +n\eta (1+3w^{2})^{n-1}(1+3w)\rho _{0}^{2n-1})=0.
\end{equation}%
There are the two standard solutions, $w=-1$ and $\rho _{0}=0$, but the
additional factor gives us another family of solutions, with

\begin{equation}
\rho _{0}^{2n-1}=-\frac{\kappa }{n\eta (1+3w^{2})^{n-1}(1+3w)}.
\end{equation}%
In the case of EMSG, when we choose $n=1$, this condition reduces to 
\begin{equation}
\rho _{0}=-\frac{\kappa }{\eta (1+3w)}  \label{desfam1}
\end{equation}%
which gives us a constant density, exponentially expanding solution for
every equation of state, $w,$ excluding $w=-\frac{1}{3},$ for an appropriate
sign of $\eta $. The existence of this extra de Sitter solution is
reminiscent of its appearance in GR cosmologies with bulk viscosity \cite%
{bulk}

This unusual situation suggests that we investigate the stability of these $%
n=1$ solutions. We consider a linear perturbation about the constant density
solution by writing 
\begin{align}
\rho =& \rho _{0}(1+\delta ), \\
H=& H_{0}(1+\epsilon ),
\end{align}%
The perturbed continuity equation is then given by 
\begin{equation}
\rho _{0}\dot{\delta}=-3(1+w)H_{0}(1+\epsilon )\rho _{0}(1+\delta )\frac{%
\kappa +\eta \rho _{0}(1+\delta )(1+3w)}{\kappa +\eta \rho _{0}(1+\delta
)(3w^{2}+8w+1)},
\end{equation}

If we use the expression for $\rho _{0}$ given in \eqref{desfam1}, we can
reduce this to 
\begin{equation}
\dot{\delta}=-3(1+w)(1+3w)H_{0}(1+\epsilon )(1+\delta )\delta \frac{1}{%
(3w^{2}+5w)(1+\frac{3w^{2}+8w+1}{3w^{2}+5w}\delta )}.
\end{equation}

From the perturbation of the modified Friedmann equation we find that $%
\epsilon \sim \delta $ which means that after expanding to first order in $%
\delta $, we have 
\begin{equation}
\dot{\delta}=-3H_{0}\delta \frac{(1+w)(1+3w)}{(3w+5)w},
\end{equation}%
so small perturbations evolve as 
\begin{equation}
\delta \propto \exp \left( {-3H_{0}}\frac{(1+w)(1+3w)}{(3w+5)w}\right)
\label{pert}
\end{equation}

\begin{figure}[th]
\centering
\includegraphics{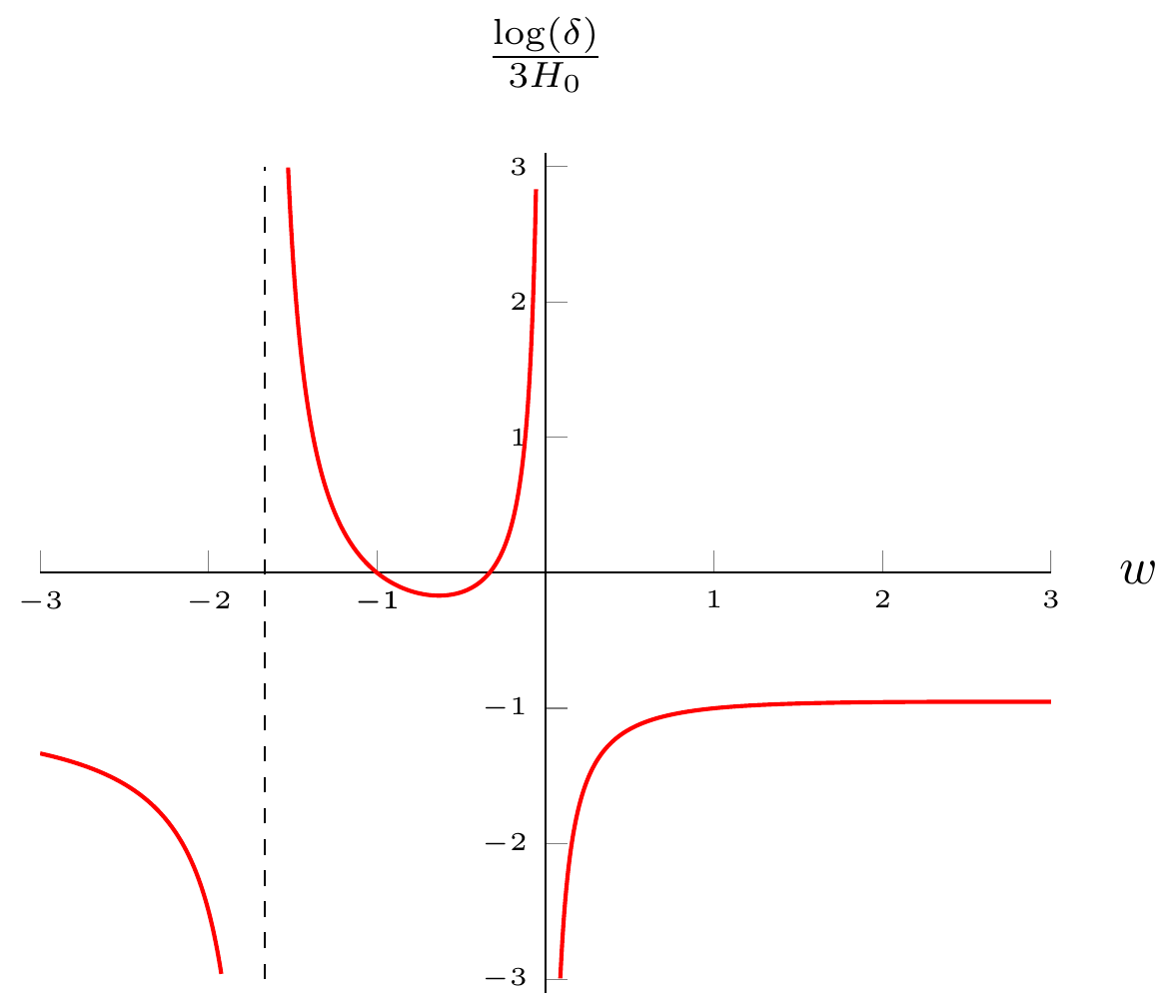} 
\caption{The plot shows the value of the exponent, and hence the
stability of the solutions, in \eqref{pert} (where we have divided by $3H_{0}
$). The asymptotes are found at $w=-\frac{5}{3}$ and $w=0$, whilst the
zeroes are at $w=-1$ and $w=-\frac{1}{3}$. The solutions will be stable for
values of $w$ where the graph is negative, and unstable otherwise.} \label%
{pertplot}
\end{figure}

The exponent in \eqref{pert} is plotted in Figure~\ref{pertplot}, where we
can see that these de Sitter-like solutions are indeed stable for a wide
range of $w$ values. This gives us an exponentially expanding universe for
(almost) any equation of state as long as we set the density to the correct
constant value. In particular, these solutions will be stable for $w<-\frac{5%
}{3}$, $-1<w<-\frac{1}{3}$ and $w>0$, and unstable for $-\frac{5}{3}<w<-1$
and $\frac{-1}{3}<w<0$. It is also the case that, depending on the sign of
the parameter $\eta $, some of these solutions will be unphysical, as they
require negative energy density. For $\eta <0$, there will be no physical
solutions for $w<-\frac{1}{3}$, whilst for $\eta >0$ there will be no
solutions for $w>-\frac{1}{3}$.

\FloatBarrier

\subsection{Early times: the bounce and high-density limits}

Examining the modified Friedmann equation \eqref{EMSGFriedmann} in the case $%
k\geq 0$, we can see that as the left-hand side of the equation is a sum of
positive terms, we must have 
\begin{equation}
\Lambda +\kappa \rho +\eta \rho ^{2}A(1,w)\geq 0,
\end{equation}

which can be split into two cases, for $\eta A(1,w)<0$ and $\eta A(1,w)>0$,
respectively. The first case occurs for 
\begin{align}
\eta <0\quad \text{and}& \quad \{w<\alpha _{-}\ \text{or}\ w>\alpha _{+}\},
\\
\eta >0\quad \text{and}& \quad \{\alpha _{-}<w<\alpha _{+}\},
\end{align}

where 
\begin{equation}
\alpha _{\pm }=-\frac{-4\pm \sqrt{13}}{3}
\end{equation}%
are the roots of

\begin{equation}
A(1,w)\equiv 3w^{2}+8w+1=0.
\end{equation}

In this case we have a maximum possible density given by 
\begin{equation}
\rho _{max}=\frac{\kappa }{2A(1,w)\eta }\left( -1+\sqrt{1-\frac{4\eta
\Lambda A(1,w)}{\kappa ^{2}}}\right) ,
\end{equation}%
indicating that a bounce occurs in this case, avoiding an initial
singularity. In the second case, where $\eta A(1,w)>0,$ there is no bounce
and no maximum energy density.

We now consider the solutions when $k=0$ in the high-density limit, where we
assume the correction terms dominate over the $\rho $ and $\Lambda $ terms.
We consider the case of general $n$, and find an analytic solution. The
Friedmann and acceleration equations reduce to 
\begin{align}
\left( \frac{\dot{a}}{a}\right) ^{2}=& \frac{\eta }{3}\rho ^{2n}A(n,w), \\
\frac{\ddot{a}}{a}=& -\frac{\eta }{3}\rho ^{2n}B(n,w).
\end{align}

From these, we can eliminate $\rho $ to find 
\begin{equation}
\left( \frac{\dot{a}}{a}\right) ^{2}+\frac{A(n,w)}{B(n,w)}\frac{\ddot{a}}{a}%
=0.
\end{equation}

which has the solution 
\begin{equation}
a(t)=D[(A+B)t-C]^{\frac{A}{A+B}}
\end{equation}

where $C$ and $D$ are new constants of integration. We can then solve for
the density: 
\begin{equation}
\rho (t)=\left( \frac{3A}{\eta }\right) ^{\frac{1}{2n}}((A+B)t-C))^{-\frac{1%
}{n}}.
\end{equation}%
This solution is real (and thus not unphysical) only if $A(n,w)/\eta $ is
positive. In the case of EMSG, this condition reduces to the requirement
that $\eta $ and $3w^{2}+8w+1$ must have the same sign. So, the two regions
where this solution exists are, 
\begin{align}
\eta >0\quad \text{and}& \quad \{w<\alpha _{-}\ \text{or}\ w>\alpha _{+}\},
\\
\eta <0\quad \text{and}& \quad \{\alpha _{-}<w<\alpha _{+}\},
\end{align}

where 
\begin{equation}
\alpha _{\pm }=-\frac{-4\pm \sqrt{13}}{3}  \label{alpha}
\end{equation}%
are the roots of $3w^{2}+8w+1$. These are complementary to the conditions
for the bounce to occur, as previously discussed. This is as we would
expect, with the high-density approximation failing at a maximum density, as
in the case of a bounce.

\section{Anisotropic Cosmology}

\label{sec:anisocossec}

There are several ways of introducing anisotropy into our cosmological
models. We will consider the simplest generalisation of FLRW, in which we
have a flat, spatially homogeneous universe, with anisotropic scale factors.
This is the Bianchi type $I$ universe, with metric given by \cite{LL} 
\begin{equation}
ds^{2}=-dt^{2}+a^{2}(t)dx^{2}+b^{2}(t)dy^{2}+c^{2}(t)dz^{2},  \label{aniso}
\end{equation}%
where $a(t)$, $b(t)$ and $c(t)$ are the expansion scale factors in the $x,y$
and $z$ directions, respectively.

Assuming that the energy-momentum tensor takes the form of a perfect fluid
with principal pressures, $p_{1}$, $p_{2}$ and $p_{3}$, so $L_{m}=\frac{1}{3}%
(p_{1}+p_{2}+p_{3})$, we can derive the field equations for Bianchi I \
universes in our higher-order matter theories:

\begin{equation}  \label{Ia}
\frac{\dot{a}\dot{b}}{ab}+\frac{\dot{b}\dot{c}}{bc}+\frac{\dot{c}\dot{a}}{ca}%
=\kappa \rho +\frac{\eta }{6}(\rho ^{2}+\sum_{i=1}^{3}p_{i}^{2})^{n-1}\left[%
(6n-3)\rho ^{2}+8n\rho
\sum_{i=1}^{3}p_{i}+2n(\sum_{i=1}^{3}p_{i})^{2}-3\sum_{i=1}^{3}p_{i}^{2}%
\right],
\end{equation}

\begin{equation}  \label{Ib}
\frac{\dot{b}\dot{c}}{bc}+\frac{\ddot{b}}{b}+\frac{\ddot{c}}{c}=-\kappa
p_{1}+\frac{\eta }{6}(\rho ^{2}+\sum_{i=1}^{3}p_{i}^{2})^{n-1}\left[2n(\rho
+p_{1}-p_{2}-p_{3})(2p_{1}-p_{2}-p_{3})-3\sum_{i=1}^{3}p_{i}^{2}\right],
\end{equation}

\begin{equation}  \label{Ic}
\frac{\dot{c}\dot{a}}{ca}+\frac{\ddot{c}}{c}+\frac{\ddot{a}}{a}=-\kappa
p_{2}+\frac{\eta }{6}(\rho ^{2}+\sum_{i=1}^{3}p_{i}^{2})^{n-1}\left[2n(\rho
+p_{2}-p_{3}-p_{1})(2p_{2}-p_{3}-p_{1})-3\sum_{i=1}^{3}p_{i}^{2}\right],
\end{equation}

\begin{equation}  \label{Id}
\frac{\dot{a}\dot{b}}{ab}+\frac{\ddot{a}}{a}+\frac{\ddot{b}}{b}=-\kappa
p_{3}+\frac{\eta }{6}(\rho ^{2}+\sum_{i=1}^{3}p_{i}^{2})^{n-1}\left[2n(\rho
+p_{3}-p_{1}-p_{2})(2p_{3}-p_{1}-p_{2})-3\sum_{i=1}^{3}p_{i}^{2}\right].
\end{equation}

In the case of an isotropic pressure fluid ($p_{1}=p_{2}=p_{3}=p$):

\begin{equation}
\frac{\dot{a}\dot{b}}{ab}+\frac{\dot{b}\dot{c}}{bc}+\frac{\dot{c}\dot{a}}{ca}%
=\kappa \rho +\frac{\eta }{2}(\rho ^{2}+3p^{2})^{n-1}((2n-1)\rho ^{2}+8n\rho
p+(6n-3)p^{2}),
\end{equation}

\begin{equation}
\frac{\dot{b}\dot{c}}{bc}+\frac{\ddot{b}}{b}+\frac{\ddot{c}}{c}=-\kappa p-%
\frac{\eta }{2}(\rho ^{2}+3p^{2})^{n-1}3p^{2},
\end{equation}

\begin{equation}
\frac{\dot{c}\dot{a}}{ca}+\frac{\ddot{c}}{c}+\frac{\ddot{a}}{a}=-\kappa p-%
\frac{\eta }{2}(\rho ^{2}+3p^{2})^{n-1}3p^{2},
\end{equation}

\begin{equation}
\frac{\dot{a}\dot{b}}{ab}+\frac{\ddot{a}}{a}+\frac{\ddot{b}}{b}=-\kappa p-%
\frac{\eta }{2}(\rho ^{2}+3p^{2})^{n-1}3p^{2}.
\end{equation}

The first of these is the generalised Friedmann equation.

Qualitatively, we expect that the higher-order density and pressure terms
will dominate at early times to modify or remove (depending on the sign of $%
\eta $) the initial singularity when $n>1/2$, but will have negligible
effects at late times, when the dynamics will approach the flat isotropic
FLRW model. At early times, we know that in GR the singularity will be
anisotropic and dominated by shear anisotropy whenever $-\rho /3<p<\rho $.
In order to determine the dominant effects as $t\rightarrow 0$ we will
simplify to the case of isotropic perfect fluid pressures ($%
p_{1}=p_{2}=p_{3}=w\rho $). Now, we determine the dependence of the
highest-order matter terms on the scale factors, $a,b$ and $c$ from the
generalisation of the conservation equation \eqref{frttcontb} with an
anisotropic metric \eqref{aniso}. For the case with general $n$, this is

\begin{equation}
\dot{\rho}=-\ \left( \frac{\dot{a}}{a}+\frac{\dot{b}}{b}+\frac{\dot{c}}{c}%
\right) \rho (1+w)\left[ \frac{\kappa +\eta \rho ^{2n-1}n(1+3w)}{\kappa
+2\eta \rho ^{2n-1}A(n,w)}\right] ,
\end{equation}%
and so the behaviour of the density is just

\begin{equation}
\rho \propto (abc)^{-\Gamma },
\end{equation}
where

\begin{equation}
\Gamma (n,w)=(1+w)\left[ \frac{\kappa +\eta \rho ^{2n-1}n(1+3w)}{\kappa
+2\eta \rho ^{2n-1}A(n,w)}\right] .  \label{gamma}
\end{equation}%
The higher-order density terms will dominate the evolution at early times
when $n>1/2$ and we see that, in these cases, $\Gamma $ is independent of $%
\rho $ and $\eta $ as $\rho \rightarrow \infty ,$since in this limit,

\begin{equation}
\Gamma (n,w)\rightarrow \frac{n(1+3w)(1+w)}{2A(n,w)}.
\end{equation}

In the cosmology obtained by setting $n=1$ in \eqref{Ia}-\eqref{Id} we will
have domination by the nonlinear matter terms, which will drive the
expansion towards isotropy as $t\rightarrow 0$ if $\rho ^{2}$ diverges
faster than $(abc)^{-2}$ as $abc\rightarrow 0$. Thus, the condition for an
isotropic initial singularity in $n=1$ theories is that $\Gamma (1,w)>2,$ or

\begin{equation}
\frac{\ (1+3w)(1+w)}{2A(1,w)}>2
\end{equation}

When this condition holds as $t\rightarrow 0,$ the dynamics will approach
the flat FLRW metric with

\begin{equation}
a(t)\propto b(t)\propto c(t)\propto t^{2/\Gamma (1,w)}.
\end{equation}

\bigskip When $\Gamma (1,w)<2,$ the dynamics will approach the vacuum Kasner
metric with

\begin{eqnarray}
(a,b,c) &=&(t^{q_{1}},t^{q_{2}},t^{q_{3}}), \\
\sum_{i=1}^{3}q_{i} &=&\sum_{i=1}^{3}q_{i}^{2}=1.
\end{eqnarray}

This condition simplifies to four cases:

\begin{center}
\begin{tabular}{|c|c|}
\hline
$w>0$ & anisotropic singularity \\ 
$\alpha _{+}<w<0$ & isotropic singularity \\ 
$\alpha _{-}<w<\alpha _{+}$ & anisotropic singularity \\ 
$w<\alpha _{-}$ & isotropic singularity \\ \hline
\end{tabular}
\end{center}

Here, the constants $\alpha _{+}$ and $\alpha _{-}$ take the values
determined earlier in \eqref{alpha}).

In general, for arbitrary $n$, the higher-order correction terms on the
right-hand side of the field equations \eqref{fieldeq} are proportional to $%
\eta \rho ^{2n}$ when $p=w\rho ,$ and so the condition for an isotropic
singularity as $t\rightarrow 0$ becomes

\begin{equation}
\Gamma (n,w)>2n,
\end{equation}%
and the dynamics approach

\begin{equation}
a(t)\propto b(t)\propto c(t)\propto t^{2/\Gamma (n,w)}.
\end{equation}

The case for general $n$ and $w$ is problematic to simplify succinctly due
to the exponential dependence on $n$. However, we can consider specific
physically relevant equations of state individually.

For dark energy ($w=-1$) and curvature ($w=-\frac{1}{3}$) `fluids', we find
that $\Gamma (n)=0$, for all $n$, and so the condition for an isotropic
initial singularity will depend only on whether $n$ itself is positive or
negative.

For $w=0$, dust, we find 
\begin{equation}
\Gamma (n,0)=\frac{n}{2n-1},
\end{equation}
for $n\neq \frac{1}{2}$. which leads to isotropy only when $\frac{1}{2}<n<%
\frac{3}{4}$.

For radiation, $w=\frac{1}{3}$, an isotropic singularity will occur if 
\begin{equation}
\frac{n}{(\frac{4}{3})^{n-1}(2n-\frac{1}{2})}>2n,
\end{equation}

whilst for $w=1$ we find that the condition for isotropy is

\begin{equation}
\frac{n}{4^{n-1}(2n-\frac{1}{2})}>2n.
\end{equation}

In both of the latter cases, we require $n\neq\frac{1}{4}$.

A similar effect will occur in more general anisotropic universes, like
those of Bianchi type $\text{VII}_{h}$ or IX, which are the most general
containing open and closed FLRW models, respectively. In type IX, the
higher-order matter terms will prevent the occurrence of chaotic behaviour
with $w<1$ fluids on approach to an initial or final singularity in a $\mathbf{T}^{2n} $ theory when $n>1$. Thus we see that in these theories the
general cosmological behaviour on approach to an initial and (in type IX
universes) final singularity is expected to be isotropic in the wide range of
cases we have determined, when $\Gamma (n,w)>2n$. This simplifying effect of
adding higher-order effects can also be found in the study of other
modifications to GR, for example those produced by the addition of quadratic 
$R_{ab}R^{ab}$  terms to the gravitational Lagrangian, \cite{bmidd, bmidd2}.
These also render isotropic singularities stable for normal matter (unlike
in GR). If $T_{ab}$ is not a perfect fluid but has anisotropic terms (for
example, because of a magnetic field or free streaming gravitons \cite{skew}%
) they will add higher-order anisotropic stresses.

\section{Conclusions}

We have considered a class of theories which generalise general relativity
by adding higher-order terms of the form $(T^{\mu \nu }T_{\mu \nu })^{n}$ to
the matter Lagrangian, in contrast to theories which add higher-order
curvature terms to the Einstein-Hilbert Lagrangians, as in $f(R)$ gravity
theories. The family of theories which lead phenomenologically to
higher-order matter contributions to the classical gravitation field
equations of the sort studied here includes loop quantum gravity, and bulk
viscous fluids, $k$-essence, or brane-world cosmologies in GR. This
generalisation of the matter stresses is expected to create changes in the
evolution of simple cosmological models at times when the density or
pressure is high but to recover the predictions of general relativistic
cosmology at late times in ever-expanding universes where the density is
small. However, we find that there is a richer structure of behaviour if we
generalise GR by adding arbitrary powers of the scalar square of the
energy-momentum tensor to the action. In particular, we find a range of
exact solutions for isotropic universes, discuss their behaviours with
reference to the early- and late-time evolution, accelerated expansion, and
the occurrence or avoidance of singularities. Finally, we discuss extensions
to the simplest anisotropic cosmologies and delineated the situations where
the higher-order matter terms will dominate over the anisotropic stresses on
approach to cosmological singularities. This leads to a situation where the
general cosmological solutions of the field equations for our higher-order
matter theories are seen to contain isotropically expanding universes, in
complete contrast to the situation in general relativistic cosmologies. In
future work we will discuss the observational consequences of higher-order
stresses for astrophysics.


\textbf{Acknowledgements}

The authors are supported by the Science and Technology Facilities Council (STFC) of the UK and would like to thank T. Harko
and O. Akarsu for helpful comments.

The authors would also like to thank Sebastian Bahamonde, Mihai Marciu and Prabir Rudra for their helpful comments.

\end{document}